\documentclass[%
reprint,
superscriptaddress,
 amsmath,amssymb,
 aps,
]{revtex4-2}

\usepackage{graphicx}% Include figure files
\usepackage{dcolumn}% Align table columns on decimal point
\usepackage{bm}% bold math
\usepackage{textcomp}
\usepackage{hyperref}
\usepackage{siunitx}
\usepackage{mathtools}
\usepackage[thinc]{esdiff}
\usepackage{nicefrac}

\usepackage{hyperref}
\hypersetup{
    colorlinks=true,
    linkcolor=blue,    
    urlcolor=blue,
    citecolor=blue
    }

\begin{document}

%\preprint{APS/123-QED}

\title{Cluster dynamics in macroscopic photoactive particles}

\author{Sára Lévay} \email{slevay@unav.es}
\affiliation{Departamento de Física y Matemática Aplicada, Facultad de Ciencias, Universidad de Navarra, E-31080 Pamplona, Spain}
\author{Axel Katona}
\affiliation{Departamento de Física y Matemática Aplicada, Facultad de Ciencias, Universidad de Navarra, E-31080 Pamplona, Spain}
\author{Hartmut Löwen}
\affiliation{Institut für Theoretische Physik II: Weiche Materie,
Heinrich-Heine-Universit{\"a}t Düsseldorf, D-40225 Düsseldorf, Germany}
\author{Raúl Cruz Hidalgo}
\affiliation{Departamento de Física y Matemática Aplicada, Facultad de Ciencias, Universidad de Navarra, E-31080 Pamplona, Spain}
\author{Iker Zuriguel}
\affiliation{Departamento de Física y Matemática Aplicada, Facultad de Ciencias, Universidad de Navarra, E-31080 Pamplona, Spain}

\date{\today}% It is always \today, today,

\begin{abstract}
We present an experimental study on the collective behavior of macroscopic self-propelled particles that are externally excited by light. This property allows testing the system response to the excitation intensity in a very versatile manner. We discover that for low excitation intensities, clustering at the boundaries is always present, even when this is prevented by implementing flower-shaped confining walls. For high excitation intensities, however, clusters are dissolved more or less easily depending on their size. Then, a thorough analysis of the cluster dynamics allows us to depict a phase diagram depending on the number of agents in the arena and the excitation intensity. To explain this, we introduce a simple kinetic model where cluster evolution is governed by a balance between adsorption and desorption processes. Interestingly, this simple model is able to reproduce the phase space observed experimentally.

\end{abstract}

\maketitle

%\tableofcontents

%%%%%%%%%%%%%%%%%%%%%%%%%%%%%%%%%%%%%%%%%%%%%%%

\emph{Introduction ---}\label{sec:intro}
Active matter refers to systems made up of many interacting self-propelled agents that convert energy into mechanical motion, exemplifying out-of-equilibrium systems~\cite{baconnier2025self}. This field has gained significant attention over the past few decades due to its broad applications in physics, chemistry, biology, medicine, and robotics. In its broader conception, examples of active matter systems include microrobots~\cite{palagi2016structured, hu2018small, xu2019millimeter, cui2019nanomagnetic, miskin2020electronically}, colloidal particles~\cite{wensink2008aggregation,buttinoni2013dynamical,palacci2013living,lozano2016phototaxis,vutukuri2020light}, bacterial systems~\cite{arlt2018painting,be2019statistical}, vibrated granular ~\cite{martinez2009enhanced, scholz2018rotating, bar2020self, soni2020phases, mohammadi2020dynamics}, robotic swarms~\cite{NISHINARI2006132,rubenstein2014programmable,oh2017bio}, animal groups~\cite{ariel2015locust, ginelli2015intermittent}, and pedestrians. 
Despite the variety of constituents and interactions within these systems, they share common properties at the group level. Accordingly, several models have been developed to describe the emergence of collective behavior, illustrating the Aristotelian concept that the whole is greater than the sum of the parts.

The energy that drives the motion of active particles can originate directly from the particles themselves, such as in bird flocks, or through an external source, as it occurs for microswimmers. Often, this motion is a response to a stimulus, a phenomenon known as taxis. Based on the nature of the driving force, we can differentiate between various types of taxis: chemotaxis~\cite{liebchen2018synthetic}, thermotaxis~\cite{mori1995neural}, viscotaxis~\cite{liebchen2018viscotaxis}, and magnetotaxis~\cite{hu2018small, xu2019millimeter, cui2019nanomagnetic, sun2019magnetic}.
%The magnetotactic behavior has some promising applications in drug delivery, as well as in the removal of blockages from blood vessels.
When the stimulus is created with light, we have phototactic or photoactive behavior~\cite{palacci2013living, lozano2016phototaxis, palagi2016structured, arlt2018painting, miskin2020electronically, vutukuri2020light}. Importantly, all of these systems involve particles whose size lies in the range of microns. 

Here we focus on a new scenario in which the active particles responding to an external stimulus (light) are inertial, polar, and macroscopic (in the scale of centimeters)~\cite{cucaslight}. This photoactive granular matter is especially interesting as the interactions occur solely through physical contact, facilitating the control and reproduction of experimental conditions. By contrast, this can be challenging in systems that involve hydrodynamic or social interactions among agents. Moreover, by using a programmable illumination panel, we have complete control over the activity of the agents, enabling us to adjust their behavior in both space and time. While this idea of external control of the activity of the agents is commonly applied to microscopic particles~\cite{arlt2018painting, vutukuri2020light}, it is pioneering in the context of macroscopic ones. To our knowledge, this approach has only been applied in phototactic wheeled robots~\cite{mijalkov2016engineering, leyman2018phototactic, rey2023light} where the dynamics are clearly different. 

Our agents are an evolved version of the HEXBUG nano\textregistered~\cite{hexbugs}, particles that are characterized by displaying a rather persistent directed motion and have been utilized in studies related to clogging, sorting, traffic jams, robotic superstructures, and resetting~\cite{patterson2017clogging, deblais2018boundaries, barois2019characterization, barois2020sorting, leoni2020surfing, boudet2021collections, baconnier2022selective, patterson2023fundamental, altshuler2024environmental, xi2024emergent}. An interesting aspect of active particles with persistent directed motion is their tendency to accumulate at the walls ~\cite{araujo2023steering}, hence leading to clustering. Clusters exhibit rich physics, depending on factors like the particle shape, driving mechanism, strength of external forcing, and interaction between agents. These factors determine the properties of the clusters, such as the size distribution, stability, and internal structure~\cite{wensink2008aggregation,martinez2009enhanced,maloney2020clustering, thapa2024emergent, becton2024dynamic, caprini2024dynamical}. Understanding the mechanisms beyond clustering in active systems is essential not only for theoretical advancements but also for applications such as programmable materials, biomedical applications, robotic swarms, and environmental sensing. In this Letter, we focus on the clustering behavior of photoactive particles under homogeneous illumination investigating the effect of particle activity and particle density.

\begin{figure}[t]
\includegraphics[width=0.425\textwidth]{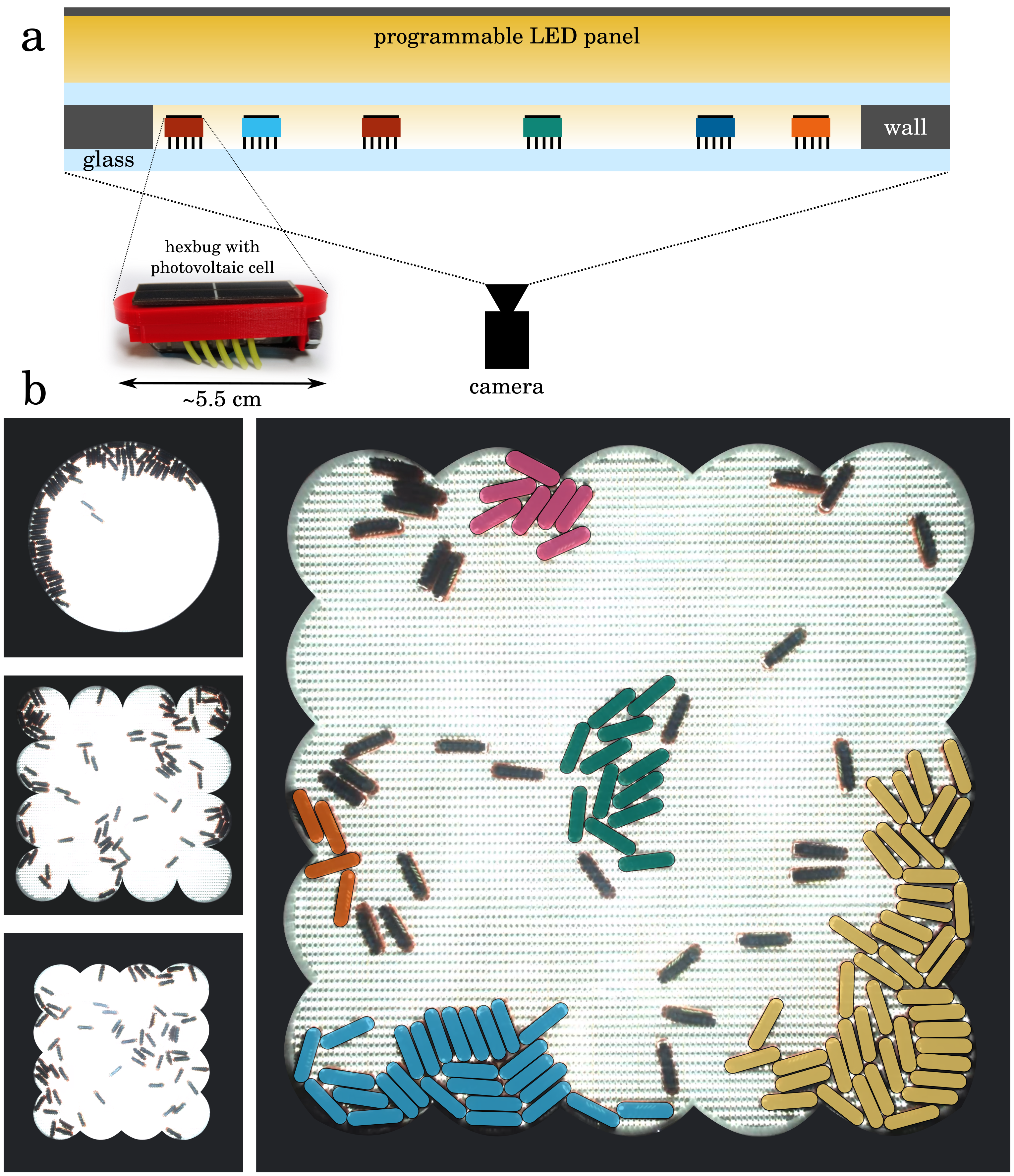}
\caption{\label{Fig:setup}(a) Scheme of the experimental setup and picture of the photoactive bug. (b) Snapshots of experiments with different walls. The largest snapshot corresponds to the arena used in the main manuscript, where different colors are used to identify clusters (see SM Video 1).}
\end{figure}

\emph{Experimental setup and methods ---}\label{sec:exp}
Our photosensitive, self-propelled particles have ten asymmetric soft rubber legs and a vibrating motor, which give rise to a directed movement with velocities depending on the vibration frequency. Originally, the motor was powered by a battery but this was replaced by a 3D-printed plastic cap holding a photovoltaic cell [see Fig.~\ref{Fig:setup} (a)]. In this way, the particles (\num{5.5}~cm~long and \num{1.5}~cm~wide) can only move when illuminated. By adjusting the illumination intensity, we can change the vibration frequency and thus control the velocity of the particles in the range of roughly \numrange{8}{14}~cm/s.

Our experimental setup is presented in Fig.~\ref{Fig:setup} and in Supplemental Material (SM)~\cite{SM}. The agents move on a horizontal glass sheet measuring \numproduct{80x80}~cm$^2$~and are enclosed by 3D-printed flower-shaped plastic walls.  Various wall structures were tested, as depicted in Fig.~\ref{Fig:setup}(b) and the effect of which is discussed in the SM. %~\cite{SM}.
The results described in this Letter were obtained using the wall structure in the right panel of Fig.~\ref{Fig:setup}(b), which was specifically designed to guide the particles inward and away from the boundaries~\cite{kumar2014flocking}. Above the arena, another glass sheet is placed at a height such that there is no contact with the bugs during their free motion but it prevents tumbling in strong collisions. In some instances, particularly in scenarios with high density and activity, a few particles (less than 5\% of the total population) were observed to tumble, hence stopping their motion. The illumination panel (\numproduct{80x80}~cm$^2$) is made of LED lines mounted on an aluminum plate. The lines are controlled by ESP32 microcontrollers, making the panel fully programmable and allowing for the application of illumination fields with temporal and spatial gradients. The imaging system (an Imaging Source DFK-37BUX250 camera) is positioned beneath the setup to record videos at $30$ fps in order to track the motion of the particles. Before studying the collective behavior of the agents, we characterized them individually to have detailed information about the illumination intensity dependence of their motion. For details, we refer the reader to the SM. %~\cite{SM}.

We used five different population sizes ($N_T=$\numlist{40;60;80;100;120}, the total number of particles in the arena) and seven different illumination intensity levels. The later is quantified by the power ($P$) that an individual bug receives through the solar cell; then, the higher the power is, the higher the bug activity is. For each population size and activity level, we conducted five independent six-minute experiments. Before each experiment, the particles were evenly distributed in rows between the two halves of the arena, facing the center (see the beginning of SM videos). Then, the light is switched on, and the particles start moving. Rapidly, driven by collisions and wall interactions, small clusters start forming either at the central part of the arena or at the boundaries [green, orange, and pink in Fig.~\ref{Fig:setup}~(b)]. Most of them just dissolved after a few seconds, but others (especially those in the walls) grew and became more stable [blue and yellow in Fig.~\ref{Fig:setup}~(b)]. In what follows, we present an analysis of the dynamics of these clusters which, in our Letter, are defined as groups of at least four particles that remain in contact for at least one second. See the SM for a detailed description of the cluster detection method. 

\begin{figure}[b]
\includegraphics[width=0.49\textwidth]{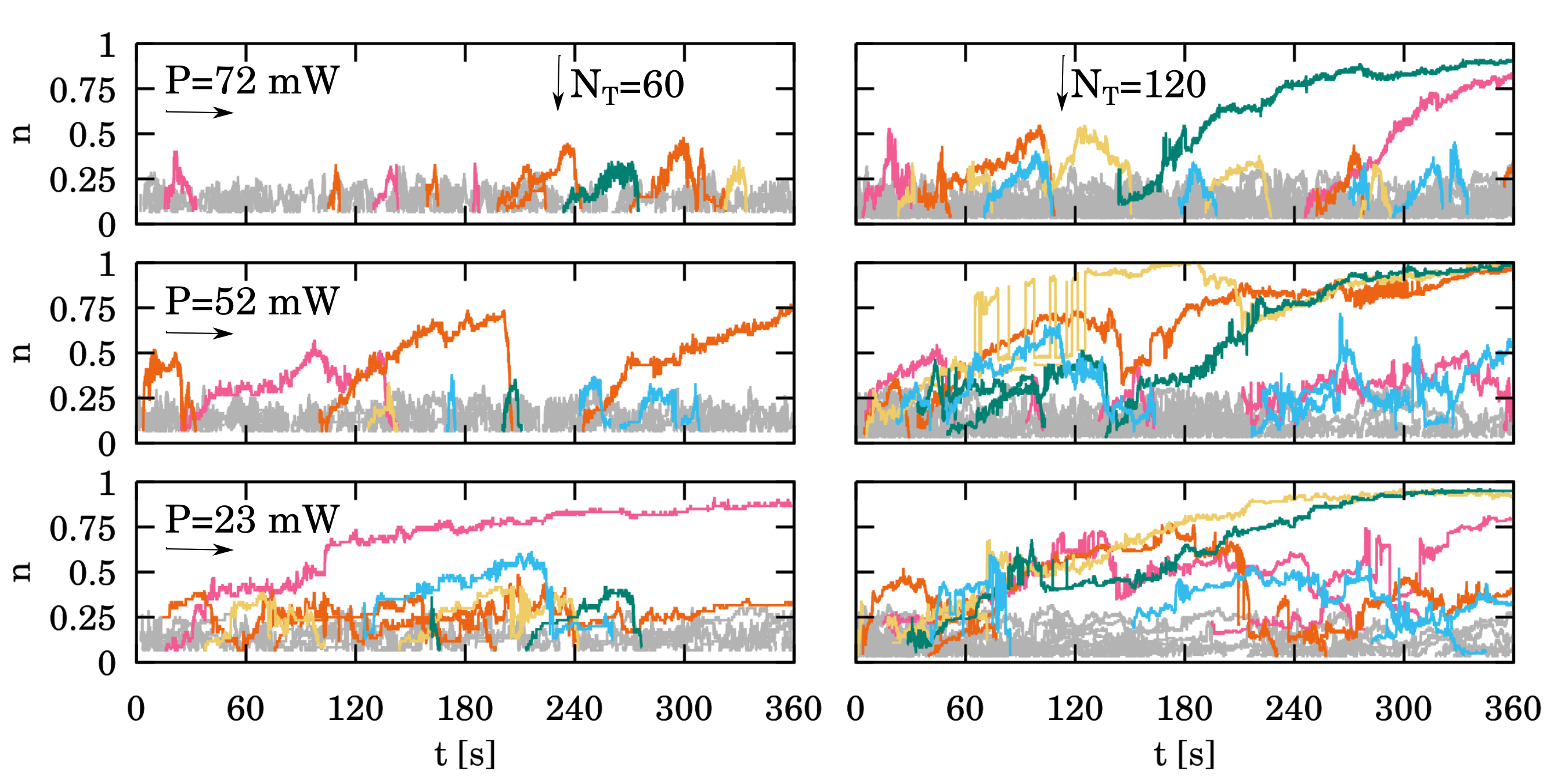}
\caption{\label{Fig:timeevol}Time evolution of cluster size, normalized by the total number of agents in the arena ($n=N/N_T$), for different activity levels and populations over six-minute experiments. Clusters reaching at least one-third of the total population are color-coded, while smaller clusters are shown in gray. The three activity levels correspond to the highest, middle, and lowest illumination intensities used in the experiments.}
\end{figure}

\emph{Results ---}\label{sec:res}
Our findings indicate that cluster size and stability increase as particle activity decreases and population size grows. This trend is illustrated in Fig.~\ref{Fig:timeevol}, which presents the temporal evolution of clusters for three illumination intensities ($P$) and two population sizes ($N_T$). For clarity, only clusters reaching at least one-third of the total population ($n=N/N_T\geq1/3$) are highlighted, with distinct colors assigned to each experimental realization. At high activity levels and low population sizes (top left panel), clusters form and rapidly disintegrate, as seen in SM Video 3~\cite{SM}. As activity decreases and population size increases, larger and more stable clusters emerge. Some clusters exhibit gradual growth, eventually incorporating a substantial fraction of the system’s particles (SM Video 1). Additionally, certain large clusters persist throughout the entire experiment, fluctuating in size as active particles can attach or detach (SM Video 2). This behavior suggests a transition from unstable to stable clustering as activity decreases and population size increases.

\begin{figure}[b]
\includegraphics[width=0.45\textwidth]{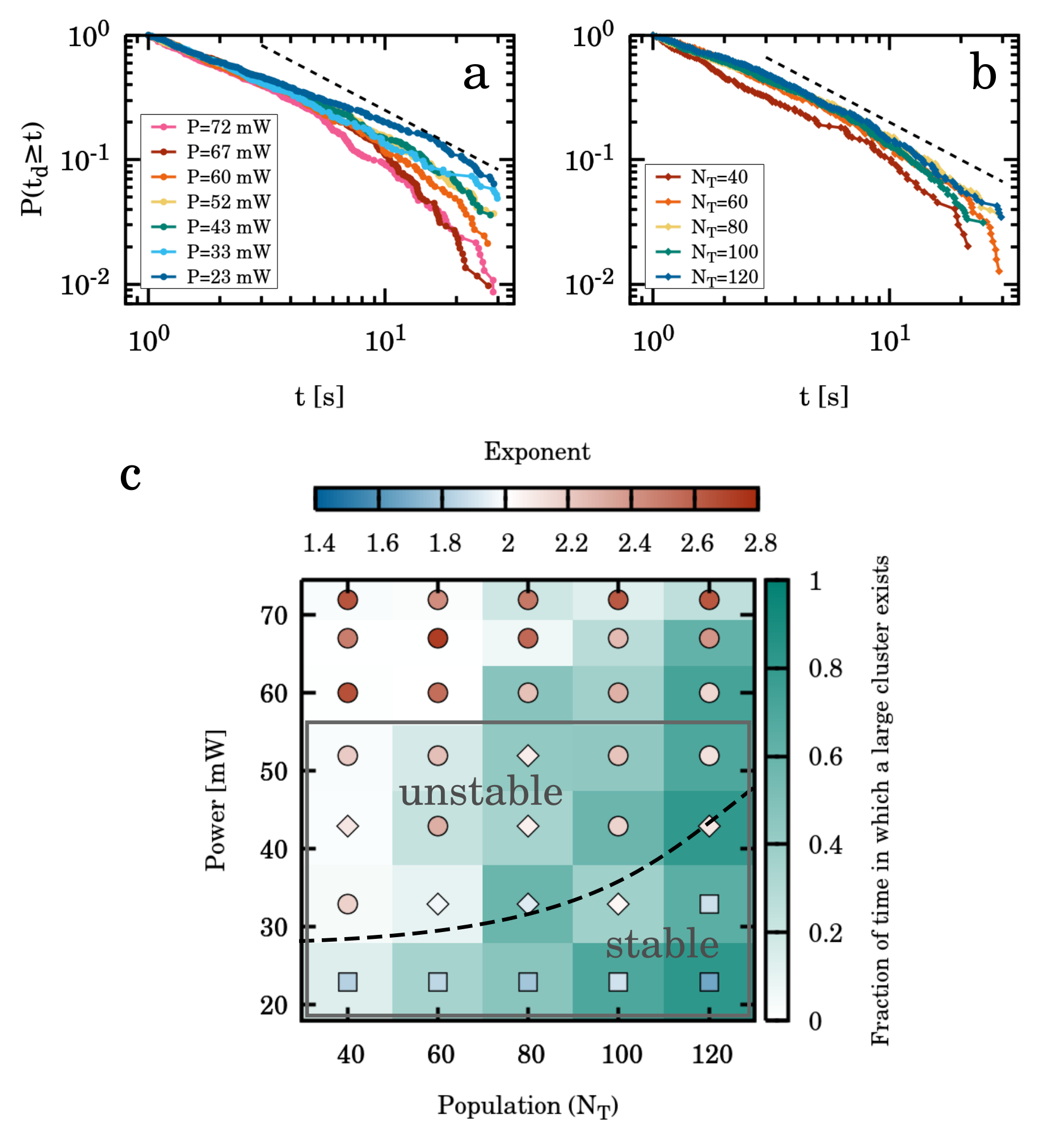}
\caption{\label{Fig:transition}(a) Survival function of cluster duration, $P(t_d \geq t)$,  for $N_T=80$ at different illumination intensities. (b)~Same function for varying population sizes at ($P=52$~mW). The dashed line represents a power law with exponent $1$ ($\alpha=2$ for the PDF). (c) Phase diagram of cluster stability: circles ($\alpha>2$, unstable), squares ($\alpha<2$, stable), and diamonds ($\alpha=2 \pm 0.1$, transition). The color of the symbol indicates the exact value of $\alpha$ as shown in the top color bar. The green shading of each internal cell shows the fraction of time in which we have at least one cluster larger than one-third of the population. The dashed black line is a guide for the eye separating the stable and unstable phases. The gray rectangle indicates the area shown later in the phase diagram obtained with the model [Fig.~\ref{Fig:model}(c)].}
\end{figure}

To analyze this transition, we examine cluster statistics, focusing on cluster duration $t_d$~\cite{transient}. Figures \ref{Fig:transition}(a) and \ref{Fig:transition}(b) presents the survival function [$P(t_d \geq t)$, also called the complementary cumulative distribution function] of the cluster duration, which represents the probability of a cluster persisting beyond a given duration $t$. Panel (a) shows the distributions for $N_T=80$ and different illumination intensities, while panel (b) compares different population sizes at an intermediate intensity ($P=52$~mW). Our results indicate a power-law-like distribution, where the probability density function (PDF) follows $\text{PDF}(t_d)\propto{t_d}^{-\alpha}$. Consequently, the survival function behaves as $P(t_d \geq t)=\int_t^\infty dt' \text{PDF}(t') \propto t^{-(\alpha-1)}$. We find that, for high activity and small populations, $\alpha > 2$, while for low activity and large populations, $\alpha \leq 2$. The latter case corresponds to diverging average cluster durations, which aligns with the presence of stable clusters persisting throughout the experiment. Fig.~\ref{Fig:transition}(c) presents a phase diagram based on the fitted exponents, illustrating that low illumination and large populations favor stable clustering (bottom right, blue symbols), whereas high illumination and small populations result in unstable clusters (top-left, red symbols). To further assess cluster stability, we measure the fraction of time in which at least one large cluster (larger than one-third of the total population) exists. These results, also displayed in Fig.~\ref{Fig:transition}(c) confirm the trends observed by the exponent analysis.

\emph{Kinetic model ---}\label{sec:model}
Inspired by Refs.~\cite{peruani2010cluster, ginot2018aggregation}, we propose a simple kinetic model aiming to capture the formation, growth, and destruction of a cluster within a "gas" of free-moving particles. The idea is that the evolution of the cluster size $N$ is governed by a balance between adsorption and desorption processes.
First, we assume that each collision between a particle and a cluster results in particle entrapment. Therefore, the adsorption rate is approximated by $k_a = v \rho_g L$, where $L$ is the typical length of the cluster, $v$ is the velocity of a free particle, and $\rho_g$ represents the density of particles in the gas phase. Then, if the cluster area is $A_c=L^2$ and $A_c=N A_p/\Phi_c$, where $A_p$ is the area of a single particle, and $\Phi_c$ is the packing fraction inside the cluster, it results that 
$k_a = v \left[ (N_T-N) / A_T \right] \sqrt{ N A_p / \Phi_c}$.
For the desorption process, we propose that particles trapped in a cluster experience a diffusive motion induced by their intrinsic vibration and collisions with their neighbors. Then, we assume that only surface particles depart from the cluster with a characteristic time $\tau_d = A_{\text{acc}}/D$, where $A_{\text{acc}}=A_p/\Phi_c$ is the accessible area for a surface particle and $D$ is the diffusion coefficient of the hexbugs within the clusters. Thus, the zero-order desorption rate is $k_d = ( D \Phi_c / A_p )$. Summing up, the cluster size $N$ follows a simple kinetic adsorption-desorption equation:
\begin{equation}
\label{eq:kinetic1}
    \diff{N}{t} = v  \frac{\left(N_T-N\right)}{A_T} \sqrt{\frac{N A_p}{\Phi_c}} - \frac{D \Phi_c}{A_p},
\end{equation}

After the arrangement of Eq.~\eqref{eq:kinetic1}, we obtain that the kinetic evolution of $n = N/N_T$ (the rescaled number of particles in a cluster) follows
\begin{equation}
\label{eq:kinetic2}
   \diff{n}{\tau} = 
   (1-n)n^{1/2} - \kappa,
\end{equation}
with
\begin{equation}
\label{eq:kappa}
\kappa= \frac{\Phi_c^{3/2} A_T}{A_p^{3/2}} \cdot \frac{1}{N_T^{3/2}} \cdot \frac{D}{v},
\end{equation}
and $\tau=c_a t $ (for details see the SM). Thus, the kinetic control parameter $\kappa$ mainly depends on three terms: the first one is a group of constants depending on the geometry of the bugs, size of the arena, etc.; the second depends on the population size ($N_T$); and the third depends on $D/v$, which in turn is related to the activity of the bugs, i.e., the light intensity.

\begin{figure}[t] 
\includegraphics[width=0.45\textwidth]{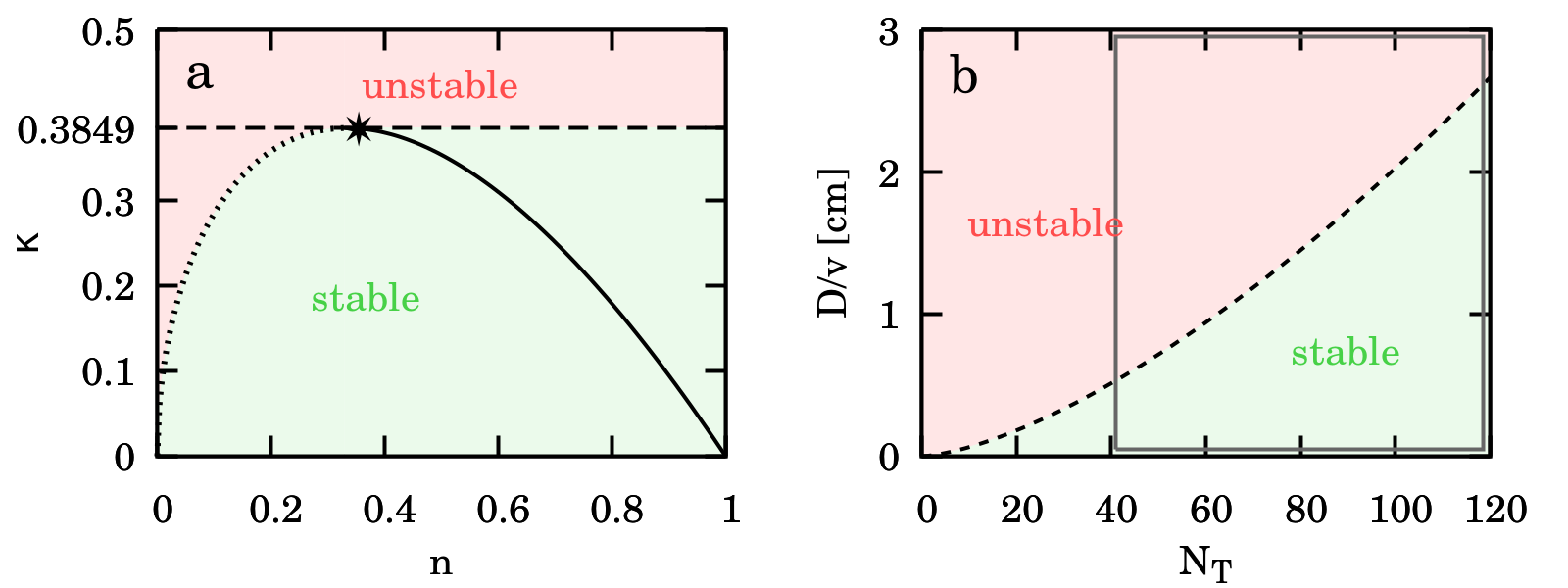}
\caption{(a) Phase diagram of the kinetic model in terms of the variables of Eq.~\eqref{eq:kinetic2}, $\kappa$ and $n$. The green region corresponds to stable clustering, whereas the red one corresponds to cluster dissolution. The dotted and continuous lines correspond to $\kappa = (1-n)n^{1/2}$. The dotted line is the minimal cluster size necessary to observe clustering, and the continuous line represents the final cluster size as $\tau\to \infty$. The star indicates the crossover between these two trends. The dashed line indicates the critical value $\kappa_{\text{crit}}=0.3849$, above which clusters are always unstable. (b) Phase diagram in terms of the total population $N_T$ and particle activity $D/v$, calculated from Eqs.~\eqref{eq:kinetic2} and \eqref{eq:kappa}, which is directly comparable to the experimental phase diagram presented in Fig.~\ref{Fig:transition}(c). For high populations and low activities, stable clusters are present, while for low populations and high activities, the clusters are unstable. The gray rectangle shows the estimated overlap with the experimental phase diagram.}
\label{Fig:model} 
\end{figure}

From Eq.~\eqref{eq:kinetic2} we can obtain the phase diagram of the model in terms of the variables $n$ and $\kappa$ [Fig.~\ref{Fig:model}(a)]. By drawing $\kappa=(1-n)n^{1/2}$ (dotted and continuous lines) we differentiate two regions: above the line $\diff{n}{\tau}<0$, so the cluster size reduces, and below the line, $\diff{n}{\tau}>0$, so the cluster size augments. Interestingly, the model predicts a critical value of $\kappa_{\text{crit}}=2\sqrt{3}/9 \approx 0.3849$ above which 
clusters always disassociate. For lower values one can have stable clusters (green region) provided that they have a certain critical initial size (dotted line). For a detailed analysis of the phase diagram, we refer the reader to SM~\cite{SM}.

In order to make a comparison of this model with the experimental results, in Fig.~\ref{Fig:model}(b) we represent the phase diagram in terms of $N_T$, the total population, and $D/v$, a variable that should be related to the activity of the particles, i.e., the light power (P). Analogous to the experiments [Fig.~\ref{Fig:transition}(c)], for small $N_T$ and high values of $D/v$, clusters disassociate, while for large $N_T$ and low values of $D/v$, clusters are stable. Importantly, while the populations of the experimental and theoretical phase diagrams are directly comparable, the correspondence between $P$ and $D/v$ is not straightforward. However, based on the phase boundaries, one can establish a relation between these variables. The corresponding parts of the phase diagrams are shown by the gray rectangles in Figs.~\ref{Fig:transition}(c) and \ref{Fig:model}(b). Also, for $N_T=80$ the model predicts the transition at $D/v=1.45$~cm, a magnitude that corresponds, for example, to $v=10$~cm/s and $D {\approx} 14.5~\text{cm}^2/\text{s}$, figures that seem rather reasonable.

\begin{figure}[b!] 
\includegraphics[width=0.45\textwidth]{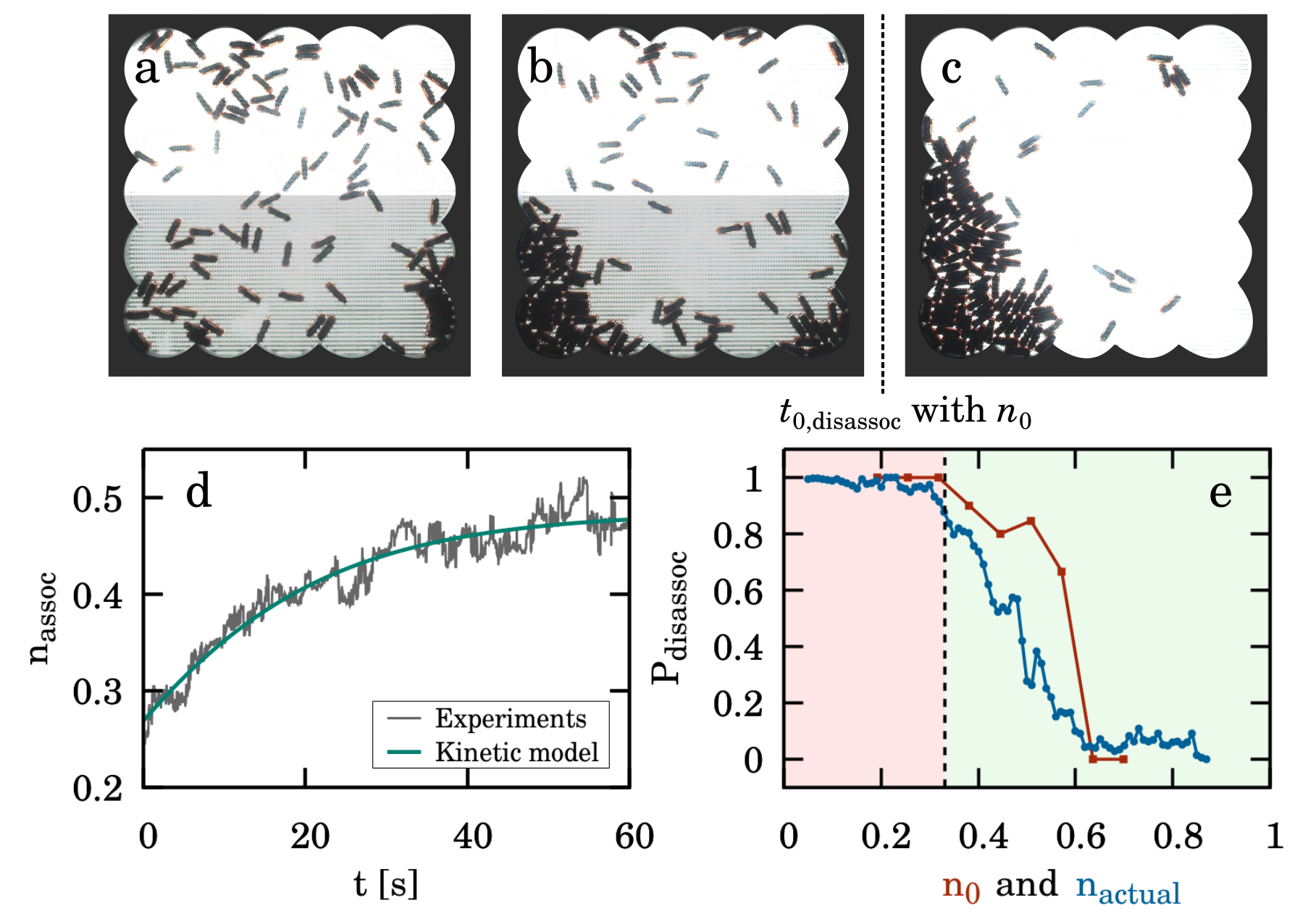}
\caption{Experiments implemented to test the cluster association and disassociation dynamics predicted by the model. We start with a homogeneously distributed sample of $N_T=100$ and we apply the highest intensity of light to one half of the arena, and the lowest to the other (a). Usually, after less than one minute, a cluster develops in the darker half (b). Then at $t_{0,\text{disassoc}}$, we switch to homogeneous illumination with the highest intensity (c) and analyze the cluster dynamics. (d) Growing phase of the clusters, averaged over $50$ independent realizations, and the fitting with the expression obtained from the kinetic model. (e) Probability of disassociation as a function of the initial cluster size $n_0$ and as a function of the cluster size at each time step ($n_{\text{actual}}$, sampled every $0.033$~s). The dashed line shows the prediction of the kinetic model for the critical cluster size above which clusters are stable.}
\label{Fig:expmod} 
\end{figure}

Aiming for a further test of the appropriateness of the model to represent the experimental behavior and taking advantage of the versatility of our setup, we conducted a controlled experiment to analyze, isolatedly, the cluster growth and disassociation. Using a population of $N_T=100$ bugs, we initially applied the maximum and minimum illumination intensities to opposite halves of the arena. This promoted cluster formation in the low-activity region [Figs.~\ref{Fig:expmod}(a) and \ref{Fig:expmod}(b)]. Then, once a sufficiently large cluster was formed [panel (b)], we switched to uniform high-intensity illumination across the arena [panel (c)] and analyzed the cluster's (disassociation) dynamics for six minutes. The time of switching to the highest intensity is denoted by $t_{0,\text{disassoc}}$, and the relative cluster size $n_0$. This protocol was repeated $50$ times to obtain good statistics. 

In the following, we analyze the cluster growth and dissolution separately. Taking into account only the association term of Eq.~\eqref{eq:kinetic1}, we obtain $n_{\text{assoc}}(t) = (N_T / N) \tanh^2{\left[\frac{1}{2} \sqrt{N_T} \left( Ct + c\right)\right]}$ for the cluster growth, where $C= ( v A_p^{1/2} / A_T \Phi_C^{1/2})$ and $c$ is a parameter depending on the initial conditions. In panel (d) we take the average growth curve of $50$ clusters and fit it accordingly, with $N_T$, $C$, and $c$ as parameters. As it can be seen, the functionality predicted by the association part of the model is recovered experimentally. Surprisingly at first, the best fitting value for $N_T$ was about 50, half of the number of hexbugs in the arena. The origin of this low value is attributed to two main causes: (1) a rather large number of hexbugs keep moving in the brighter side of the arena, so they do not contribute to the cluster growing, and (2) most of the times two clusters are formed in the darker side of the arena, so the smallest one does not contribute to the growing of the largest. Beyond this, from the fitted value of $C$, we can estimate the free particle velocity, obtaining $v \approx 6.8$~cm/s, which is in good agreement with the actual experimental values. 

Moving to the disassociation process, in panel (e) we present the probability of cluster disassociation (after six minutes) as a function of the initial cluster size ($n_0$, red). Importantly, the value of $n_0$ at which the probability starts to decay nicely resembles the value offered by the model if we consider the highest value of $D/v$ that allows the development of stable clusters ($0.33$, dashed line). As an alternative method to estimate the disassociation probability, we consider the cluster size at all times as initial sizes ($n_{\text{actual}}$, blue) and determine the probability of dissolution at the end of the experiment. Clearly, the results of this method resemble the previous ones, confirming the predictive character of our model.

In this Letter, we investigate cluster dynamics using a novel class of photoactive macroscopic particles. We identify a stable clustering phase at low particle activities and high population sizes, whereas at high activities and low population sizes, clusters rapidly dissolve. This behavior is captured by a kinetic model incorporating particle adsorption and desorption rates. The model reveals an undiscovered dependence of the cluster stability on the initial cluster size that is posteriorly corroborated by new experiments in which we take advantage of the external control of the particles' activity. Beyond the interest of these results, we believe that this Letter may inspire new investigations in which the macroscopic active agents are driven at will by smartly controlling the supplied energy fields. In this way, the implementation of alternating intensity fields (SM Video 4) and traveling waves (SM Video 5) can be seen as natural extensions of this Letter.

\emph{Acknowledgments ---}
We especially acknowledge Luis Fernando Urrea for the technical help and Lucia Sastre for helping with the preparation of the particles. This project has received funding from the European Union’s Horizon 2020 research and innovation programme under the Marie Skłodowska-Curie Grant Agreement No. 101067363, named PhotoActive, and the Spanish Government through grant No. PID2023-146422NB-I00 supported by MICIU/AEI/10.13039/501100011033. A. K. acknowledges the Asociación de Amigos, Universidad de Navarra, for a grant.

\emph{Data availability ---}
The data that support the findings of this article are openly available \cite{data, code}.

\bibliography{main}% Produces the bibliography via BibTeX.

\begin{thebibliography}{54}
\bibitem [{tra()}]{transient}%
  \BibitemOpen
  \href@noop {} {}\bibinfo {note} {The first $2$ minutes of the $6$-minute-long experiments were removed from the analysis in order to prevent the inclusion of a transient that would depend on the initial sample preparation.}\BibitemShut {Stop}%
\end{thebibliography}

%%%%%%%%%%%%%%%% SUPPLEMENTARY

\pagebreak

\onecolumngrid
\begin{center}
  \textbf{\large Supplemental Material for ``Cluster dynamics in macroscopic photoactive particles"}\\[.2cm]
  Sára Lévay,$^{*}$ Axel Katona, Hartmut Löwen, Raúl Cruz Hidalgo, Iker Zuriguel\\[.1cm]
  ${}^*$Contact author: slevay@unav.es\\
%(Dated: \today)\\[1cm]
\end{center}

\setcounter{equation}{0}
\setcounter{figure}{0}
\setcounter{table}{0}
\setcounter{page}{1}
% change figure labels
\renewcommand{\thefigure}{S\arabic{figure}}
\renewcommand{\theequation}{S\arabic{equation}}
\renewcommand{\bibnumfmt}[1]{[S#1]}
\renewcommand{\citenumfont}[1]{S#1}
\renewcommand{\thesection}{S~\Roman{section}}

\section{Experimental setup}

\begin{figure}[b]
\includegraphics[width=0.75\textwidth]{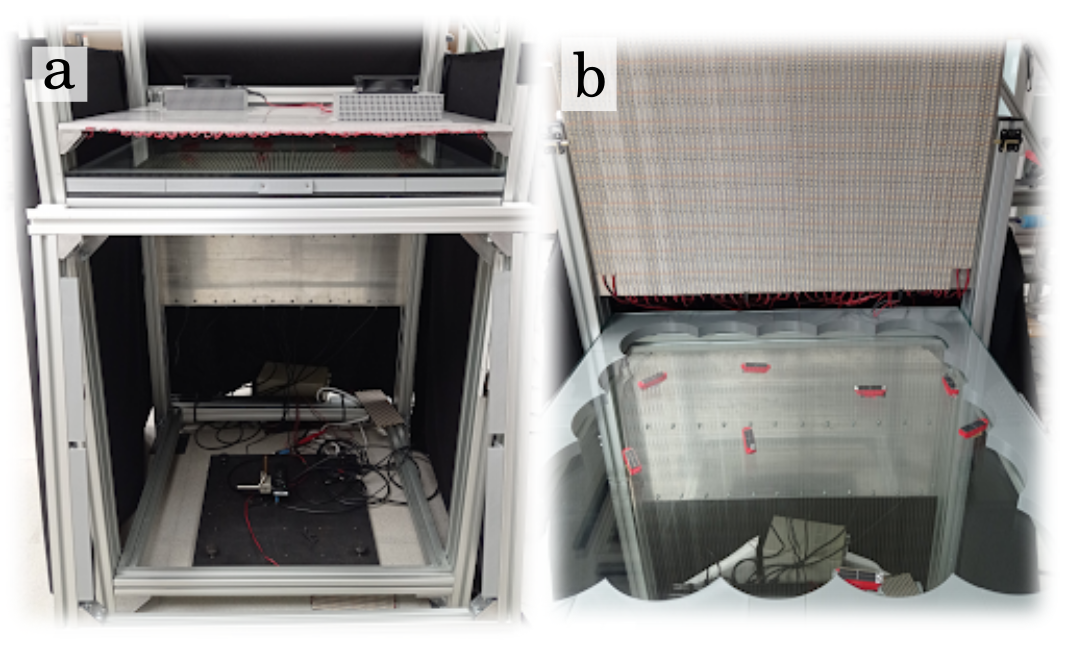}
\caption{\label{Fig:SMsetup}Photos of the experimental setup. (a): Side view. (b): Top view with the illumination panel lifted, showing a few particles and the enclosing wall structure.}
\end{figure}

Fig.~\ref{Fig:SMsetup}~(a) shows the side view of the experimental setup. The glass sheets and the illumination panel are mounted on an aluminum scaffold. The imaging system is placed beneath the experiment. The illumination panel can be easily lifted as shown in panel (b), and the upper glass sheet can be slid to manipulate the particles. Additionally, two fans are mounted on the top of the illumination panel to prevent overheating.

The illumination panel is made of LED lines, mounted on an aluminum plate, controlled by ESP32 microcontrollers. Serial port communication method was established between a computer using LabView and the microcontrollers. By this we can tune the illumination intensity of each line individually, using pulse-width modulation.

\section{Videos}
\href{https://www.youtube.com/watch?v=jQXGe59rCVs&list=PLWYWlSkkYOSN4V0nXWi6b2MPm_CT9drO5}{SM Videos}

\textbf{Video 1} shows an experiment conducted with $120$ particles under low illumination intensity ($23$~mW). The particles were initially arranged in rows, evenly distributed between the two halves of the system, facing towards the center. Smaller clusters formed and destroyed throughout the experiment; then one cluster gradually grew and became stable, ultimately incorporating around $75$\% of the particles. The experimental photo shown in Fig.~1~(b) of the main manuscript was captured during this experiment (see at 1'28'').

\textbf{Video 2} shows an experiment with $100$ particles under medium illumination intensity ($43$~mW). Throughout the experiment, clusters formed and destroyed, without any stable clustering observed. Towards the end of the video, we present a case, where two clusters are merged, resulting in the creation of a relatively large cluster that was later destroyed.

\textbf{Video 3} depicts an experiment with $60$ particles under high illumination intensity ($72$~mW). The high particle activity leads to the rapid formation and destruction of small clusters. No cluster becomes stabilized within the experimental duration.

\section{The kinetic model}

Aiming for a better understanding of the time evolution of clusters in the system, we have developed a simple kinetic model that describes the fundamental processes responsible for cluster growth and destruction. 
Note that this model is deterministic as it does not include noise or randomness of any nature. Obviously, this is not the case in real experiments but still, we believe it is useful to understand the observed behavior. 

Basically, we propose that the variation of the number of particles in a cluster is governed by a balance of adsorption and desorption of active particles. Then, to estimate the rate of particles adsorbed by the cluster, we assume that every collision of a free particle with a cluster results in particle entrapment. The collision rate depends on the relative velocity of the cluster and the single particle, the particle density in the `gas' phase ($\rho_g$), and the typical length of the cluster ($L$). The absorption rate is thus $k_a = v \rho_g L$. We assume that the relative velocity of the cluster and the single particle is mostly governed by the latter with velocity $v$. The particle concentration in the gas phase is $\rho_g=(N_T-N)/A_T$, where $N_T$ is the total number of particles in the system, $N$ is the number of particles in the cluster and $A_T$ is the total area of the arena. Then, we relate the length of the cluster with the area of the cluster by assuming $A_c=L^2$. In addition, the area of the cluster can be expressed by $A_c=N A_p/\Phi_c$, where $A_p$ is the area of a single particle and $\Phi_c$ is the packing fraction within the cluster. From this $L=\frac{N^{1/2}A_p^{1/2}}{\Phi_c^{1/2}}$ resulting in the absorption rate:
\begin{equation}
    k_a = v \cdot \frac{N_T-N}{A_T} \cdot \frac{N^{1/2}A_p^{1/2}}{\Phi_c^{1/2}} = \frac{ v A_p^{1/2} }{ \Phi_c^{1/2} A_T } (N_T-N)N^{1/2}.
\end{equation}

To estimate the rate of the departure of active particles from the cluster, we assume that the motion of the surface particles in the cluster can be described within the diffusion approximation. The rate of desorption of a particle from the cluster is $k_d = D/A_{\text{acc}}$, where $D$ is the coefficient of diffusion of particles in the cluster surface, and $A_{\text{acc}}$ is the accessible area for a cluster particle: $A_{\text{acc}}=\frac{A_p}{\Phi_c}$. This results in the desorption rate:
\begin{equation}
    k_d = \frac{D\Phi_c}{A_p}.
\end{equation}

The variation of the number of particles in a cluster is then:
\begin{equation}
    \diff{N}{t} = k_a - k_d = \frac{ v A_p^{1/2} }{ \Phi_c^{1/2} A_T } (N_T-N)N^{1/2} - \frac{D\Phi_c}{A_p}.
\end{equation}
Introducing $n=N/N_T$, as the relative cluster size, the equation takes the following form: 
\begin{equation}
    \diff{n}{t} = c_a (1-n)n^{1/2} - c_d,
\end{equation}
with $c_a = \frac{ v A_p^{1/2} N_T^{1/2} }{ \Phi_c^{1/2} A_T }$ and $c_d = \frac{D\Phi_c}{A_p N_T}$, containing fixed parameters and the two variables we change in the experiments, namely the total number of particles ($N_T$), and the illumination intensity. The model reflects the latter through $D$ and $v$.
Rescaling the time to $\tau=c_a t$, the equation turns into:
\begin{equation}
    \diff{n}{\tau} = (1-n)n^{1/2} - \kappa,
\end{equation}
where 
\begin{equation}
    \kappa = c_d/c_a = \frac{\Phi_c^{3/2} A_T}{A_p^{3/2}} \cdot \frac{1}{N_T^{3/2}} \cdot \frac{D}{v},
\end{equation}

\noindent where the constants (first term) and the parameters changed in the experiments (second and third terms) are separated.

\begin{figure}[t]
\includegraphics[width=0.75\textwidth]{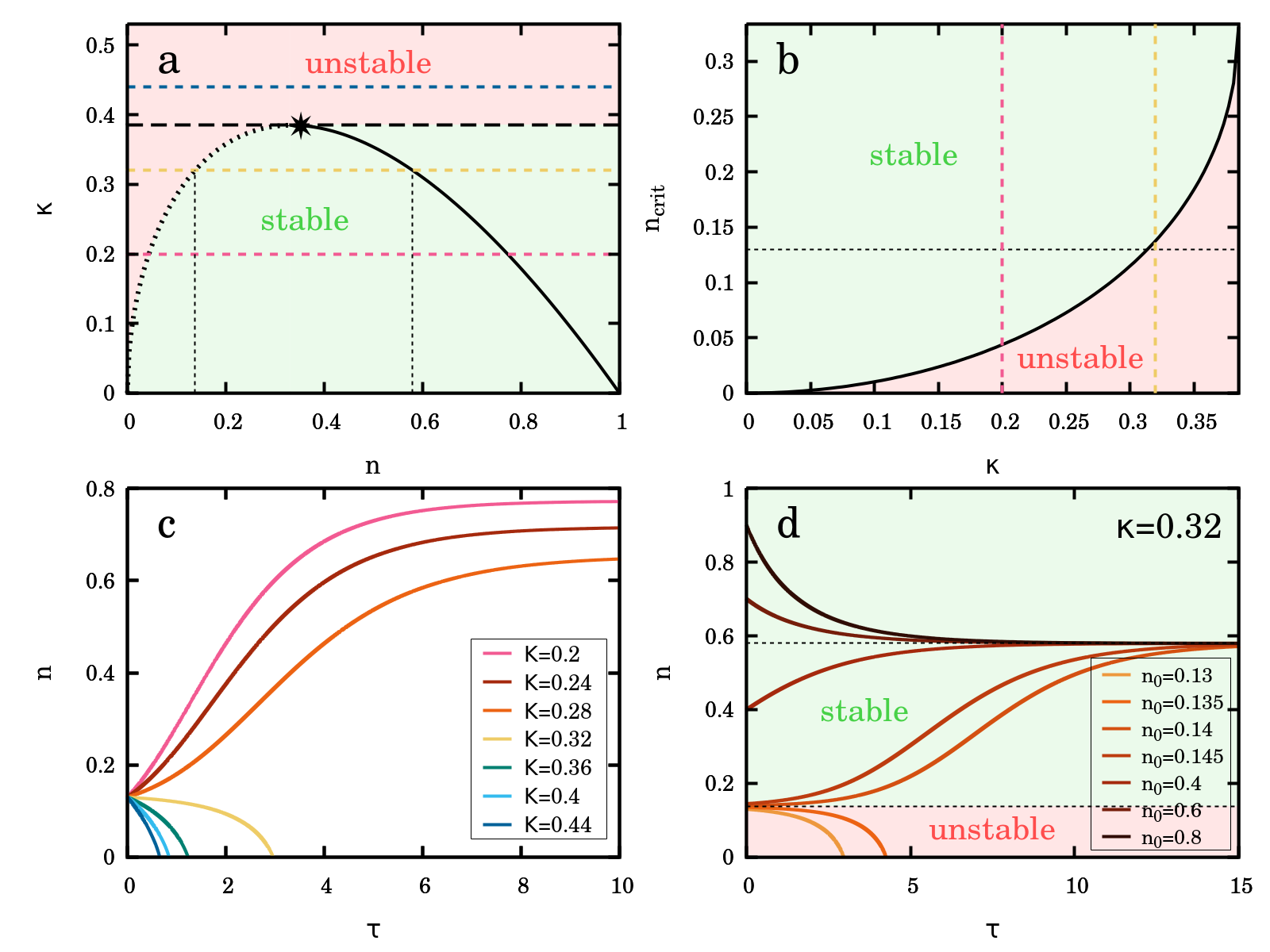}
\caption{\label{Fig:SMmodel}(a): Phase diagram identical to the one presented in Fig.~4~(c) of the main paper but with the inclusion of several reference (dashed) lines for several values of $\kappa$ that are analyzed in the other panels. (b): The critical initial cluster size, $n_{\text{crit}}$ as a function of $\kappa$. Recall that $\kappa_{\text{crit}}=0.3849$, so $n_{\text{crit}}$ diverges at this value. The yellow and pink dashed lines show the same value of $\kappa$ as in panel (a). (c): Time evolution of clusters predicted by the model for different $\kappa$ values, starting from an initial cluster size $n_{0}=0.13$. The pink, yellow, and dark blue curves correspond to the lines of the same color in panels (a-b). Note that in panel (b) there is no blue curve as it is beyond $\kappa_{\text{crit}}$. (d): Time evolution of clusters predicted by the model for $\kappa=0.32$ (yellow line in panels (a-b), starting from different initial cluster sizes ($n_0$). Below $n_{\text{crit}}=0.1377$ (light orange lines) stable clusters do not develop, while above that threshold stable clusters develop.}
\end{figure}

If $\kappa>(1-n)n^{1/2}$ (red area in Fig.~\ref{Fig:SMmodel}~(a)), thus $\diff{n}{\tau}<0$, clusters are not stable, they disassociate and disappear. Otherwise, when $\kappa<(1-n)n^{1/2}$, hence $\diff{n}{\tau}>0$, then clusters grow and stabilize (green area in Fig.~\ref{Fig:SMmodel}~(a)). This means that for values of $\kappa$ above the critical value $\kappa_{\text{crit}}= 2\sqrt{3}/9 \approx 0.3849$, clusters always decay, and for values of $\kappa$ below the critical value, clusters grow and stabilize if they start from a size larger than an initial critical cluster size, $n_{\text{crit}}$ (see dotted curve in Fig.~\ref{Fig:SMmodel}~(a)). This behavior is better understood by looking at Fig.~\ref{Fig:SMmodel}~(b), where we only represent the minimum cluster size necessary to reach stable clustering as a function of $\kappa$. Clearly, $n_{\text{crit}}$ diverges for $\kappa_{\text{crit}}=0.3849$, as for values of $\kappa$ above this threshold, clusters are not stable, no matter their size. 

Note that the parameter $\kappa$ decreases as we increase the population in the system and decrease the particle activity (thus decreasing $D/v$). This means that the kinetic model captures the same transition between unstable and stable clustering that we have seen in the experiments. Then, aiming a better understanding of the model, in Fig.~\ref{Fig:SMmodel}~(c) we show the time evolution of clusters for different values of $\kappa$, starting from an initial cluster size $n_0=0.13$ (dashed line in panel (b)). This initial cluster size allows the formation of stable clusters for values of $\kappa$ slightly below $0.32$. For cases with higher values of $\kappa$, thus lower populations and higher particle activities, the formation of stable clusters is no longer possible, the desorption term is dominating in their time evolution.

To stress the importance of initial cluster size $n_0$, Fig.~\ref{Fig:SMmodel}~(d) shows the time evolution of clusters for $\kappa=0.32$ (yellow line in panels (a) and (b)), starting from different initial sizes. One can see, that for $n_0$ values below $n_{\text{crit}}(\kappa{=}0.32)=0.1377$ (light oranges), the clusters are not stable and disappear after a relatively short time. Above this critical initial cluster size (dark oranges) clusters grow and stay stable at the final stable cluster size ($0.5797$ for $\kappa{=}0.32$). When the initial cluster size is larger than this value (dark reds), the model predicts an initial decrease in the cluster size, stabilized at this value.

\section{Influence of the wall structure on the clustering dynamics}

As it is broadly known, self-propelled particles tend to cluster at the walls of the arena as, when they collide perpendicularly, they need several seconds to change the direction of motion. This is typically the seed for the cluster generation which occurs when other particles interact with this one. An alternative scenario that may lead to the creation of a cluster is when two particles moving along the wall in opposite directions, collide and start pushing each other. As a result of this, the confining walls have significant importance in the observed development and growth of clusters. Indeed, when the arena is circular, clustering is extremely favored, occurring for almost all activities and densities. One would observe the same behavior in the case of straight walls, with particles accumulating at the corners. In order to prevent this, and redirect the particles to the center of the arena, flower-shaped geometries were ideated.

\begin{figure}[b]
\includegraphics[width=0.75\textwidth]{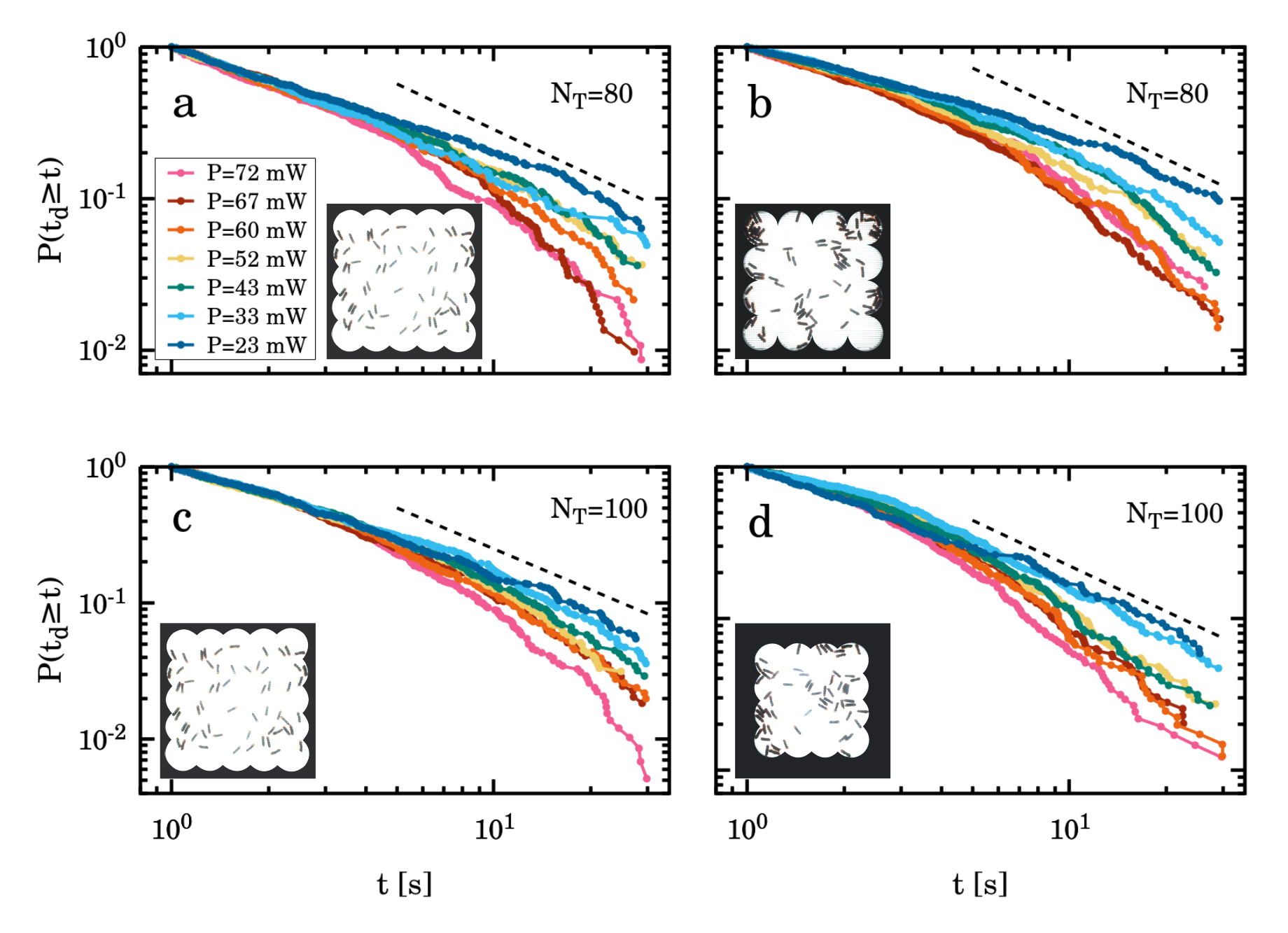}
\caption{\label{Fig:walls} Survival function of the cluster duration for different wall structures. In panels (a) and (b) we work with $N_T=80$ and compare arenas in which we increase the separation between the curved segments. In panels (c) and (d) we work with $N_T=100$ and we compare arenas with different surfaces while keeping the wall geometry. Different light intensities are implemented as indicated in the legend of panel (a).}
\end{figure}

In our experiments, the wall structure is made of small, curved segments (with a diameter of $17$~cm, which is around $3$ times the length of particles). All the experiments we present in the paper were done with the structure visible in Fig.~\ref{Fig:walls} (a,c) in which each side of the arena contains 5 of these curved segments. To test the generality of the observed behavior we experimented with two other structures. In one case (Fig.~\ref{Fig:walls} (b)) we increased the separation between the circular segments (maintaining the curvature), hence reducing the number of them that fit in each lateral side (from 5 to 4). In the other case, we kept the original structure (curvature and separation of segments) but reduced the arena size by approximately $40$\% as can be seen in panel (d).

In Fig.~\ref{Fig:walls}~(a-d) we show the cluster duration survival functions for the three different wall structures. In both cases, we repeated $5$ experiments for each illumination level. In the case of the (b) structure, we have experimented with $80$ particles, while for the (d) structure we experimented with $100$ bugs. To facilitate the comparison, in panels (a,c) we present the results of these two populations for the original wall structure. Comparing panels (a,b) we observe that the arena with 4 curved segments leads to slightly longer (more stable) clusters than the arena with 5 curved segments. This can be understood if we think that in the 4-segment arena, the area covered for each one of these circles is larger than in the 5-segment arena, a feature that could probably favor the stability of the clusters formed inside them. Similarly, reducing the arena surface but keeping the same boundary shape (panels c,d) seems to slightly increase the cluster stability; a behavior that correlates with the increment of the density of particles within the arena which, by the way, is a feature covered by the model. 

In any case, the behavior encountered for these other two wall geometries is similar to that reproduced in the main manuscript, evidencing a transition from cluster dissolution to cluster stabilization as it decreases the intensity of light. 

\section{Characterization of the motion of individual particles}

Before studying the collective behavior of the photosensitive bugs, we characterize their individual motion and obtain information about their velocity and preferred directionality. Importantly, by adding the plastic cap and the photovoltaic cell to the Hexbugs, we alter their mass distribution, which, in most cases, makes them rotate. We try to eliminate this by adding a counterbalance and small additional masses to the particles. We calibrate the position of the small masses such that the bug would be able to maintain a relatively straight trajectory under the highest level of illumination ($P{=}72$~mW), although in some cases particles show a tendency of rotation. 

\begin{figure}[b]
\includegraphics[width=0.6\textwidth]{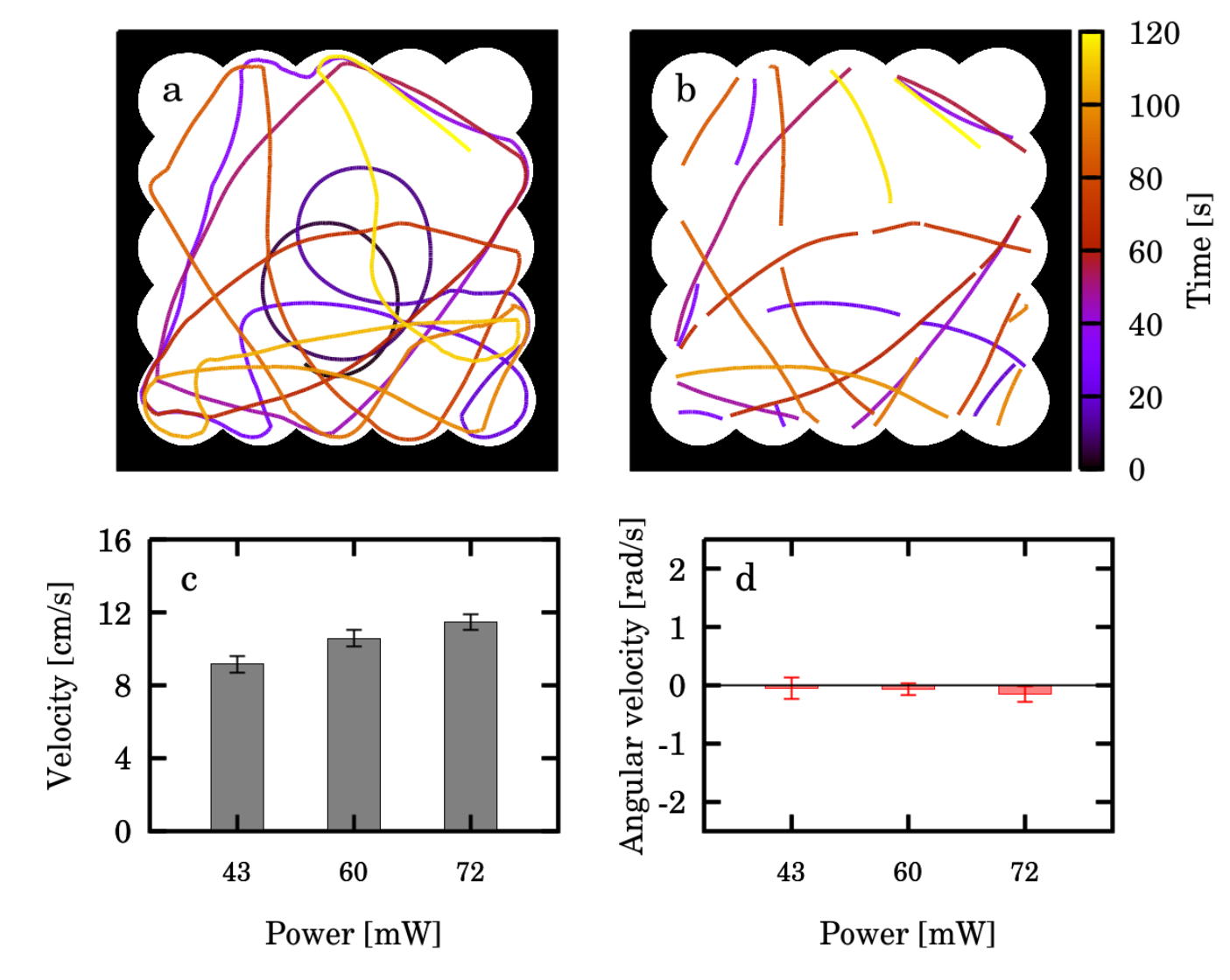}
\caption{\label{Fig:char1}Characterization of a single particle. (a): Trajectory under homogeneous illumination ($P=72$~mW) for a $2$-minute experiment. Colors represent the time. (b): Segments of the same trajectory. Particle-wall and particle-particle collisions were removed. (c): Average velocity of the same particle calculated for all the non-collisional segments of 4 repetitions of 2-minute-long experiments. (d): Average angular velocity for the same particle. As the values obtained are small and persistent over different realizations, this particle is considered to move along relatively straight lines.}
\end{figure}

In order to quantify the velocity and the persistence of the directionality of the bugs, we performed several experiments with all of the particles. In groups of $8$, the particles moved within the arena under three different homogeneous illumination intensities for two minutes. We repeated this protocol $4$ times for all groups. Then, we detect the trajectories (see Fig.~\ref{Fig:char1} (a)) and, after identifying all collisions (particle-particle and particle-wall), we obtain the segments of the trajectories between collisions (panel (b)). For all the segments we calculate the velocity and the directionality (based on the change in velocity direction) of the particles, as presented in panels (c-d). Therefore, in the plots of Fig.~\ref{Fig:char1}(c-d), the results presented for each illumination intensity were taken from 8 experimental minutes in total.

\begin{figure}[t]
\includegraphics[width=0.9\textwidth]{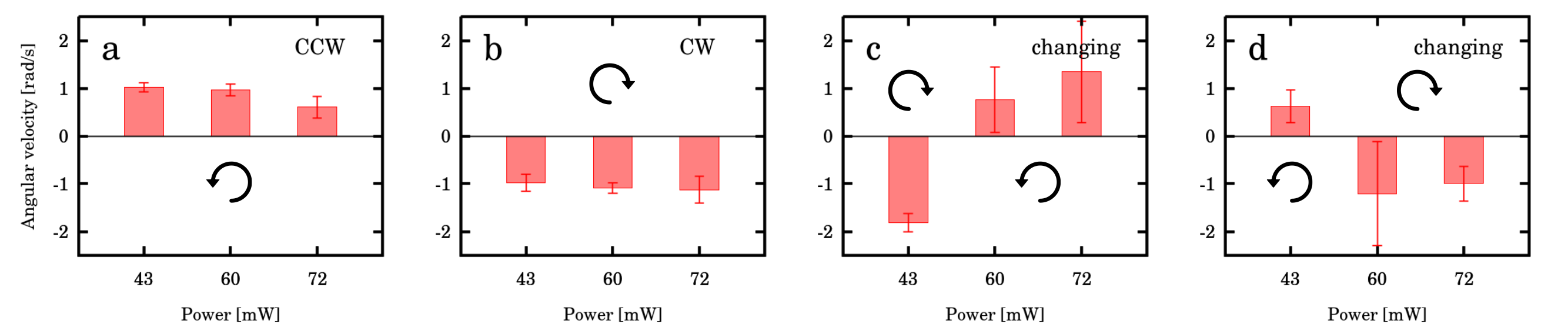}
\caption{\label{Fig:char2}Angular velocities of different particles obtained by repeating 2-minute-long experiments 4 times for three different homogeneous illumination intensities. (a): As the values are positive and relatively large (larger than 0.5~rad/s), this particle is considered to show a tendency to counterclockwise rotation. (b): Bug with a tendency to rotate clockwise. (c-d): Examples of particles with a changing rotational tendency. As we increase the illumination intensity, the direction of the rotation changes from CW to CCW or vice versa.}
\end{figure}

The particle which behavior is represented in Fig.~\ref{Fig:char1} displays a motion that is considered as straight. Indeed, segments persistently have an angular velocity smaller than 0.5~rad/s. However, this is not always the case and each agent can behave in very different manner. In Fig.~\ref{Fig:char2}, we show the angular velocities of other particles to exemplify the different types of behavior encountered among our bugs. In panels (a-b) we show particles that consistently tend to rotate clockwise or counterclockwise as evidenced by the absolute value of the angular velocity which is persistently larger than $0.5$~rad/s. Importantly, in these particles the directionality does not change with the activity level. However, there are other particles whose rotation direction changes as with the illumination intensity (panels (c-d)). 

Based on these encountered behavior, we grouped our 200 particles into different subsets: around $50\%$ of the particles move more or less straightly (Fig.~\ref{Fig:char1}), $22\%$ tend to rotate counterclockwise (Fig.~\ref{Fig:char2} (a)), another $22\%$ clockwise (Fig.~\ref{Fig:char2} (b)), and around $6\%$ of the particles change the direction of rotation as we change the illumination intensity (Fig.~\ref{Fig:char2} (c-d)). In terms of the magnitude of the velocities, we have only seen a slight variation among different particles used, clearly showing increasing speeds with increasing illumination intensity.

In the experiments presented in the manuscript, for cases with $N_T{\leq}60$, we used particles exclusively from the first group (straight motion), while for higher populations we took particles from both of the groups. In terms of the collective behavior and the cluster dynamics, we have not seen any important differences that could be related to the specific composition of the population. However, as particles collide frequently, small alterations of the mass distribution may happen, causing some particles (usually less than $10$\% of the population) to develop strong rotation (as can be seen in some of the videos). These particles were replaced by new ones during the experiments.

\section{Cluster detection method}
In order to study the cluster dynamics the particles were not tracked individually. Instead, patches of particles in contact with one another were detected with conventional image detection methods, due to the relatively high contrast between particle-occupied and empty regions. The number of particles constituting a patch is estimated from its area. For each frame of the recorded videos, we identify these patches and for further analysis, we consider only those that contain at least $4$ particles. We then create a virtual network in which the nodes are the detected patches. Two nodes are connected by a link if they are from subsequent frames and have at least $30\%$ overlap (this means that, the areas of the patches in consecutive frames overlap at least $30\%$). In this way linked parts of the network will represent the time evolution of a given patch. We apply a clustering algorithm to find the connected components in the network, with the criteria that a cluster should be at least $1$ second long. If a connected component has branches longer than $2$ seconds, we treated the branch as an individual cluster. By applying this method, properties such as duration, size, and the temporal evolution of all clusters can be quantified.

\end{document}